\documentclass[aps,nofootinbib,prd,preprint,longbibliography,preprintnumbers,showpacs]{revtex4-1}
\pdfoutput=1
\usepackage{graphicx}
\usepackage{amsmath,amssymb,mathrsfs}
\usepackage{bbm}
\usepackage{color}
\usepackage{xcolor}
\usepackage{ulem}
\usepackage{dsfont}
\usepackage{cancel}
\usepackage{dsfont}
\usepackage{epstopdf}
\usepackage{epsfig}
\usepackage{bm}
\usepackage{dcolumn}
\usepackage{hyperref}
\usepackage{enumitem}

\def\beq{\begin{eqnarray}}
\def\eeq{\end{eqnarray}}
\def\beq{\begin{equation}}
\def\eeq{\end{equation}}
\def\be{\begin{eqnarray}}
\def\ed{\end{eqnarray}}
\def\non{\nonumber}

\def\gaa{\mathrel{\raise.3ex\hbox{$>$\kern-.75em\lower1ex\hbox{$\sim$}}}}
\def\la{\mathrel{\raise.3ex\hbox{$<$\kern-.75em\lower1ex\hbox{$\sim$}}}}
\newcommand{\ba}{\begin{array}}
\newcommand{\ea}{\end{array}}
\newcommand{\besub}{\begin{subequations}}
\newcommand{\eesub}{\end{subequations}}
\newcommand{\ee}{\end{equation}}
\newcommand{\bea}{\begin{eqnarray}}
\newcommand{\eea}{\end{eqnarray}}

\begin{document}

\title{{Explaining Dijet Mass Excesses \\ in ALEPH LEP2 Four-Jet Events with 2HDMs}}

\author{A.~Arhrib$^1$, R.~Benbrik$^2$, W.~Klemm$^{3,4}$, S.~Moretti$^5$, A.~Rouchad$^2$}
\affiliation{
$^1$ Facult\'e des Sciences et Techniques, Abdelmalek Essaadi University, B.P. 416, Tangier, Morocco.\\
$^2$ MSISM Team, Facult\'e Polydisciplinaire de Safi, Sidi Bouzid, B.P. 4162,  Safi, Morocco.\\
$^{3}$ Department of Physics and Astronomy, Uppsala University, Box 516, SE-751 20 Uppsala, Sweden.\\
$^{4}$ School of Physics $\&$ Astronomy, University of Manchester, Manchester M13 9PL, UK.\\
$^5$ School of Physics and Astronomy, University of Southampton,
  Southampton, SO17 1BJ, UK.
}

\date{\today}

\begin{abstract}
{In this paper, we address the observability of four-jet signatures 
from light neutral and charged Higgs bosons at LEP2 energies in the framework of 2-Higgs Doublet Models (2HDMs). 
The main signal production channels are via $e^+ e^- \to Ah$ and $H^+ H^-$ with subsequent quark decays of such final states into four-jets. Specifically, Type-I and -III realizations of a generic 2HDM (2HDM-I and -III, respectively) are adopted to show that there exist points (under the assumption that the heavy Higgs state $H$ is SM-like) for which  $(m_h, m_A, m_{H^\pm}) \approx (80, 30, 55)$ GeV that can yield observable rates at LEP2 energies that can potentially explain the di-jet mass excesses seen recently in ALEPH data in a re-analysis of their four-jet samples, particularly so for the 2HDM-III.
}
\end{abstract}

\maketitle                  

\section{Introduction}
\label{sec:intro}
Our understanding of jet physics has been constanty expanding, whether through the adoption of different algorithms for the definition of jets \cite{Moretti:1998qx} or better emulation of QCD partonic processes and/or parton shower dynamics \cite{Ellis:1991qj}, so it is warranted to revisit the cleanest of all jet data in our possession, i.e., those produced in $e^+e^-$ annihilation from a few GeV all the way up to LEP2 energies \cite{Kluth:2003yz}. In fact, unlike the case of hadronic machines, jet production in $e^+e^-$  colliders is an
ideal laboratory for QCD studies. The reason is threefold. Firstly, hadronic
final states therein do not interfere with the leptonic
initial state. Secondly, the energy of the hadronic final
state is maximal in the laboratory frame (in the
case of symmetric $e^+e^-$ colliders, which is historically most often the case) thus allowing efficient
production and experimental study of highly energetic
hadronic systems. Thirdly, since the particles involved
in the Electro-Weak (EW) production of hadrons, i.e., electrons and positrons, are point-like, there are no parton density functions to take into account.\\

When revisiting LEP2 data of the ALEPH Collaboration in which the hadronic final state was reconstructed in terms of four-jets, the authors of Ref.~\cite{Kile:2017ccn} (see also Ref.~\cite{Kile:2017psu}) have pointed out 
an excess observed in such hadronic events. The tracks are clustered into four-jets and paired such that the mass difference between the two di-jet systems is minimised. The excess occurs in the region $M_1 +M_2 \approx 110$ GeV, where $M_i$ ($i=1,2$) are the two di-jet masses ($M_1$ is defined to contain the highest $p_T$ jet). About half of the excess is concentrated in the region $M_1 \approx 80$ GeV  and $ M_2 \approx 30$ GeV, with a local significance between $4.8\sigma$  and $5.6\sigma$, while the  other half is found for  $M_1 \approx M_2 \approx 55$ GeV, with a local significance of $4.1\sigma$ to $4.5\sigma$. These results are rather robust against changes in the QCD Monte Carlo (MC) sample, the jet-clustering algorithm and the jet-energy-rescaling method. Further, no source of systematic uncertainty was found that can explain the excess. Finally, no analogue of the excess is seen at LEP1.  (Notice that no jet-flavor tagging was enforced in the analysis.)\\

We attempt in this paper to describe such excesses as being due to 2HDM \cite{Branco:2011iw} events wherein  $e^+ e^- \to Z^{(*)}h, Ah$ and $H^+ H^-$  production takes place, with the $Z$ gauge boson and $h,A$ and $H^\pm$ Higgs bosons decaying hadronically, so as to naturally produce four-jet events. Recall that the physics spectrum of a 2HDM includes two CP-even neutral Higgs states ($h$ and $H$ with, conventionally, $m_h<m_H$), one CP-odd neutral Higgs state ($A$) and a pair of charge-conjugated charged Higgs bosons ($H^\pm$). Herein, we identify the $H$ state as the SM-like Higgs boson discovered at the Large Hadron Collider (LHC) in 2012 and we will be looking at regions of 2HDM parameter space where all other Higgs states are lighter. In fact, we will be able to identify parts of the 2HDM parameter space over which an explanation for the ALEPH excess can be found with those ameanable to LHC  investigations described in Refs.~\cite{Arhrib:2017uon,Arhrib:2017wmo,Arhrib:2016wpw}, thereby providing a compelling case for a thorough assessment of 2HDMs with both old and new collider data.\\

The paper is organised as follows. The next section describes the 2HDMs we will be dealing with. Sect.~\ref{sec:param} discusses the theoretical and experimental constraints enforced on the Higgs production processes in our theoretical scenarios as well as the surviving parameter space of the latter. Sect.~\ref{sec:results} presents our results. Finally, we conclude in Sect.~\ref{sec:summa}.

\section{The model}
\label{sec:model}

In the generic 2HDM two identical Higgs doublets $\Phi_1$ and $\Phi_2$ with hypercharge $Y=+1/2$ are introduced.
The most general $SU(2)_L\times U(1)_Y$ gauge invariant potential, with dimension four terms only, can be written as:
\begin{equation}
 \begin{aligned}
   V_{\mathrm{2HDM}}=    &\,m_{11}^2\Phi_1^\dagger\Phi_1+m_{22}^2\Phi_2^\dagger\Phi_2
    -\left[m_{12}^2\Phi_1^\dagger\Phi_2+\mathrm{h.c.}\right]
    +\tfrac{1}{2}\lambda_1(\Phi_1^\dagger \Phi_1)^2 \nonumber \\
    &+\tfrac{1}{2}\lambda_2(\Phi_2^\dagger \Phi_2)^2
    +\lambda_3(\Phi_1^\dagger \Phi_1)(\Phi_2^\dagger \Phi_2)
    +\lambda_4\left(\Phi_1^\dagger\Phi_2\right)\left(\Phi_2^\dagger\Phi_1\right)\nonumber\\
    &+\left\{
    \tfrac{1}{2}\lambda_5\left(\Phi_1^\dagger\Phi_2\right)^2 +(\lambda_6 (\Phi_1^\dagger \Phi_1)
+  \lambda_7 (\Phi_2^\dagger \Phi_2)) (\Phi_1^\dagger \Phi_2)    +\mathrm{h.c.}\right\}.
  \end{aligned}
  \label{eq.pot}
\end{equation}
By hermiticity of the potential, one finds that $\lambda_{1,2,3,4}$ are real. Further,
to guarantee CP invariance of the above scalar potential,  $m_{12}$, $\lambda_{5}$, $\lambda_6$ and 
$\lambda_7$  should  also be taken as real. 
If one asks the above potential to  respect a discrete $Z_2$ symmetry requested 
for flavor conservation, $\Phi_i \to - \Phi_i$ ($i=1,2$),  then  $\lambda_6$ and  $\lambda_7$ must vanish\footnote{In general though, we will allow for a dimension two term that softly breaks the $Z_2$ symmetry.}. 
We impose that the minimum of the scalar potential preserves the $U(1)_\text{EM}$ gauge symmetry of Electro-Magnetism (EM), such that the (pseudo)scalar fields develop the following Vacuum Expectation Values (VEVs):
\begin{align}
\langle \Phi_1 \rangle = \frac{1}{\sqrt{2}} \left( \begin{array}{c} 0 \\ v_1 \end{array} \right), \qquad \langle \Phi_2 \rangle = \frac{1}{\sqrt{2}} \left( \begin{array}{c} 0 \\ v_2 \end{array} \right).
\end{align}
The two Higgs doublets can then be expanded around the potential minimum in terms of their component fields as follows:
\begin{equation}
\Phi_1 = \left(
\begin{array}{c}
\phi_1^+\\
\frac{1}{\sqrt{2}}\left(v\,\cos\beta+\phi_1^0\right)
\end{array}
\right),
\qquad
\Phi_2 = \left(
\begin{array}{c}
\phi_2^+\\
\frac{1}{\sqrt{2}}\left(v\,\sin\beta+\phi_2^0\right)
\end{array}
\right).
\end{equation}

From the original eight scalar degrees of freedom, three Goldstone bosons ($G^\pm$ and $G$) are absorbed by the $W^\pm$ and $Z$ bosons. The remaining five degrees of freedom form the aforementioned physical Higgs states of the model: $h, H, A$ and $H^\pm$.
It is  more  convenient to express the scalar doublet fields in the 
Higgs basis~\cite{Georgi:1978ri,Branco:1999fs,Davidson:2005cw}, defined by
\begin{align}
H_1 = \left(\begin{array}{c} H_1^+ \\ H_1^0 \end{array}\right) \equiv \Phi_1 \cos\beta + \Phi_2 \sin\beta\;, \qquad H_2 = \left(\begin{array}{c} H_2^+ \\ H_2^0 \end{array}\right) \equiv -\Phi_1 \sin\beta + \Phi_2 \cos\beta\;,
\label{eq:Higgsbasis}
\end{align}
such that the VEVs of these fields are $\langle H_1^0\rangle = v/\sqrt{2}$ and $\langle H_2^0 \rangle = 0$. Thus, the scalar doublet $H_1$ possesses the same tree-level couplings to all the SM particles as the SM Higgs boson. In the Higgs basis the physical Higgs states are given by
\begin{align}
\arraycolsep=5pt
\def\arraystretch{1.3}
\left(\begin{array}{c} h \\ H \end{array}\right) = 
\begin{pmatrix}
-  \sin(\beta-\alpha) & \cos(\beta-\alpha) \\
\cos(\beta-\alpha) & \sin(\beta-\alpha)
\end{pmatrix}
\left( \begin{array}{c}
\mathrm{Re}\, (H_1^0) - v\\
\mathrm{Re}\, (H_2^0)
\end{array}
\right)
\end{align}
and 
\begin{eqnarray}
H^\pm = H_2^\pm, \qquad \qquad A =\sqrt{2} {\rm Im} (H_2^0).
\end{eqnarray}
If one of the physical Higgs states is aligned with $\mathrm{Re}(H_1^0) - v$, it obtains the tree-level couplings of a SM Higgs boson. For the light neutral Higgs state 
$h$(heavy Higgs state $H$) this occurs when $\cos(\beta-\alpha)\to 0$($\sin(\beta-\alpha)\to 0$). Thus, each case can provide a possible explanation of the $125$ GeV Higgs signal~\cite{Bernon:2015qea,Bernon:2015wef}. 

After using the two minimization conditions to eliminate 
$m_{11}$ and $m_{22}$ in terms of the $\lambda_i$'s and mixing angles together with 
  the $W^\pm$ mass to eliminate one of the VEVs as a function of $m_W$ and $\tan\beta$, 
  we are left with nine independent free parameters which can be taken as
  the four Higgs masses, $\tan\beta$,  the mixing angle $\alpha$ (or $\sin(\beta-\alpha)$), 
  $m_{12}^2$, $\lambda_6$ and $\lambda_7$.
  In the case of 2HDM with flavor conservation such as 2HDM Type-I (2HDM-I) or -II (2HDM-II)  
  one can take $\lambda_6=\lambda_7=0$, thus eliminating two further parameters.

  In our analysis, we will assume that $H$ is the SM-like Higgs with  $m_H=125$ GeV, hence 
  $m_h< 125$ GeV. This assumption will force $\cos(\beta-\alpha)\approx 1$,  
   such that the coupling $HVV$  ($V=W^\pm,Z$) is SM-like as indicated by LHC data. 

Without advocating the discrete symmetry $Z_2$ in the scalar potential and in the Yukawa Lagrangian,
both Higgs doublets can couple to leptons and quarks which could lead to Flavour Changing Neutral Currents
(FCNCs) at the tree level.  
The Yukawa interactions for quarks 
are written as
\beq
{\cal {L}}_{Y} =  \bar Q_{L} 
Y^{k}  d_{R} \Phi_k+  \bar Q_{L }  \tilde{Y}^{k}  u_{R} \tilde{\Phi}_k + {\rm h.c.}
\eeq
where the flavor indices are removed, $Q^T_L=(u_L, d_L)$ is the left-handed quark 
doublet, $Y^k$ and  $\tilde{Y}^k$ denote the $3\times 3$ 
Yukawa matrices,  $\tilde{\Phi}_k=i\sigma_2 \Phi^*_k$ and $k$ is the 
doublet number. Similar formulae could be derived for the lepton sector. Thus,  the mass matrices of quarks are linear combination of  $Y^{1}$($\tilde Y^{1}$) 
and $Y^2$($\tilde Y^{2}$) for down(up)-type quarks. Therefore, in general, the diagonalization of the fermionic mass matrices does not work for $Y^{1,2}$  and $\tilde Y^{1,2}$ simultaneously. 
As a result, tree level FCNCs appear and consequent effects lead to significant 
oscillations of $K-\bar K$, $B_q-\bar B_q$ and $D-\bar D$.  
To get naturally small FCNCs, one can use the ansatz formulated as 
$Y^{k}_{ij},\,\tilde Y^k_{ij} \propto \sqrt{m_i m_j}/v$ 
in Refs. \cite{Cheng:1987rs,Atwood:1996vj}. The associated model is called 2HDM Type-III (2HDM-III) \cite{Cheng:1987rs,Atwood:1996vj,Branco:2011iw}. After  spontaneous EW Symmetry Breaking (EWSB),  
the (pseudo)scalar couplings to fermions can be expressed as \cite{GomezBock:2005hc,Arhrib:2015maa,Benbrik:2015evd}
\begin{equation}
 \begin{aligned}
 {\cal L}^{\rm III}_{\rm Y}  &=&
\bar u_{Li} \left( \frac{\cos\alpha}{\sin\beta}\frac{m_{u_i}}{v} \delta_{ij} - \frac{\cos(\beta-\alpha)}{\sqrt{2}\sin\beta} X^u_{ij} \right) u_{Rj} h + \bar d_{Li} \left( -\frac{\sin\alpha}{\cos\beta}\frac{m_{d_i}}{v} \delta_{ij}  + \frac{\cos(\beta-\alpha)}{\sqrt{2}\cos\beta} X^d_{ij} \right) d_{Rj} h \non \\ 
&+& \bar u_{Li} \left( \frac{\sin\alpha}{\sin\beta}\frac{m_{u_i}}{v} \delta_{ij} + \frac{\sin(\beta-\alpha)}{\sqrt{2}\sin\beta} X^u_{ij} \right) u_{Rj} H + \bar d_{Li} \left(\frac{\cos\alpha}{\cos\beta}\frac{m_{d_i}}{v} \delta_{ij}
 - \frac{\sin(\beta-\alpha)}{\sqrt{2}\cos\beta} X^d_{ij} \right) d_{Rj} H \non\\
&-& i\bar u_{Li} \left(\frac{1}{\tan\beta}\frac{m_{u_i}}{v} \delta_{ij} -
\frac{X^u_{ij} }{\sqrt{2}\sin\beta}  \right) u_{Rj} A 
+ i\bar d_{Li} \left(-\tan\beta\frac{m_{d_i}}{v} \delta_{ij}
 + \frac{X^d_{ij}}{\sqrt{2}\cos\beta} \right) d_{Rj} A +{\rm h.c.}\,,
\label{eq:LhY}
 \end{aligned}
\end{equation}
with
\begin{eqnarray}
X^q_{ij}=\sqrt{m_{q_i} m_{q_j}}/v \chi^q_{ij}, \qquad  \qquad q=u,d,
\label{eq:chi-ij}
\end{eqnarray}
where the $\chi^{q}_{ij}$'s are free parameters.

We may instead assume a $Z_2$ symmetry in the Yukawa Lagrangian, leading to flavor conservation
\cite{Glashow:1976nt}.  This can generate four types of Yukawa interactions, including the Type-I 2HDM (2HDM-I), in which all fermions couple to a single Higgs doublet.
In this analysis we will consider 2HDM-I and 2HDM-III, which more readily accomodate the low masses discussed here. 
The Yukawa Lagrangian of the 2HDM-I is
\begin{eqnarray}
 - {\mathcal{L}}_{\rm Y}^I = \sum_{\psi=u,d,l} \left(\frac{m_\psi}{v} \kappa_\psi^h \bar{\psi} \psi h^0 + 
 \frac{m_\psi}{v}\kappa_\psi^H \bar{\psi} \psi H^0 
 - i \frac{m_\psi}{v} \kappa_\psi^A \bar{\psi} \gamma_5 \psi A^0 \right) + \nonumber \\
 \left(\frac{V_{ud}}{\sqrt{2} v} \bar{u} (m_u \kappa_u^A P_L +
 m_d \kappa_d^A P_R) d H^+ + \frac{ m_l \kappa_l^A}{\sqrt{2} v} \bar{\nu}_L l_R H^+ + {\rm h.c.}\right),
 \label{Yukawa-1}
\end{eqnarray}
where $\kappa_{u,d,l}^h=\cos\alpha/\sin\beta$,  $\kappa_{u,d,l}^H=\sin\alpha/\sin\beta$, 
$\kappa_{u}^A=1/\tan\beta$ and $\kappa_{d,l}^A=-1/\tan\beta$.
 
 For completeness,  recall that the couplings of neutral Higgs bosons to gauge bosons  are independent of the Yukawa textures, i.e., 
 \begin{eqnarray*}
 hVV\propto \sin(\beta-\alpha),\quad  HVV\propto \cos(\beta-\alpha),  
 \quad hAZ\propto \cos(\beta-\alpha), \quad HAZ\propto \sin(\beta-\alpha) 
 \end{eqnarray*}
and note that the couplings of the  $\gamma$ and $Z$  bosons to a pair of charged Higgses are pure gauge interactions. Aside from the Yukawa couplings, these are those intervening in our upcoming numerical analysis.  

\begin{table}[h!]
	\begin{center}
		\begin{tabular}{|c|c|c|c|}
			\hline 
			\hline Observable 		&  Experimental result & SM contribution    & Combined error at 1$\sigma$\\
\hline $\mathcal B (K\to\mu\nu) / \mathcal B(\pi\to\mu\nu)$	& $0.6357 \pm 0.0011$~\cite{Agashe:2014kda} & $ 0.6231\pm 0.0071 $& 0.0071 \\
			\hline $\overline{\mathcal B}(b\to s\gamma)_{E_\gamma>1.6\,\text{GeV}}$ & $(3.32 \pm 0.16)\times 10^{-4}$~\cite{Amhis:2016xyh}& $(3.36\pm 0.24) \times 10^{-4}$ &  $0.29\times 10^{-4}$\\
			\hline $\mathcal B (B\to\tau\nu)$									& $(1.14 \pm 0.22)\times 10^{-4}$~\cite{Amhis:2014hma}& $(0.78\pm 0.07 )\times 10^{-4}$ & $0.23\times 10^{-4}$\\
			\hline $\mathcal B (D\to\mu\nu)$ & $(3.74 \pm 0.17)\times 10^{-4}$~\cite{Agashe:2014kda,Amhis:2014hma}& $(3.94\pm 0.13)\times 10^{-4}$   & $0.21 \times 10^{-4}$\\
			\hline $\mathcal B(D_s\to\tau\nu)$ & $(5.55 \pm 0.24)\times 10^{-2}$~\cite{Agashe:2014kda,Amhis:2014hma}& $(5.17\pm 0.11) \times 10^{-2}$ & $0.26\times 10^{-2}$\\
			\hline $\mathcal B (D_s\to\mu\nu)$	& $(5.57 \pm 0.24)\times 10^{-3}$~\cite{Agashe:2014kda,Amhis:2014hma}& $(5.28\pm 0.11) \times 10^{-3}$  & $0.26\times 10^{-3}$\\
			\hline $\overline{\mathcal B} (B^0_s \to \mu^+ \mu^-)$ 				& $(2.8\pm0.7)\times10^{-9}$~\cite{Archilli:2014cla} &$(3.66\pm0.28) \times 10^{-9}$ & $0.75\times 10^{-9}$ \\
			\hline $\overline{\mathcal B} (B^0_d \to \mu^+ \mu^-)$& $(3.9\pm 1.5)\times 10^{-10}$~\cite{Archilli:2014cla}& $(1.08\pm 0.13) \times 10^{-10}$ & $1.50\times 10^{-10}$ \\
			\hline $\Delta M_s$ & $(17.757\pm 0.021)\text{ ps}^{-1}$~\cite{Amhis:2016xyh}&$(18.257\pm 1.505) \text{ ps}^{-1}$ & $1.5\text{ ps}^{-1}$\\
			\hline $\Delta M_d$ & $(0.510\pm 0.002) \text{ ps}^{-1}$~\cite{Amhis:2016xyh}	& $(0.548\pm0.075) \text{ ps}^{-1}$ & $0.075 \text{ ps}^{-1}$\\
			\hline $\Delta_0 (B\to K^*\gamma)$ 									& $(5.2\pm 2.6)\times 10^{-2}$~\cite{Agashe:2014kda}	& $(5.1\pm 1.5) \times 10^{-2}$ & $3.0\times 10^{-2}$\\
			\hline $\delta a_\mu$ & $(261\pm 80)\times 10^{-11}$~\cite{Hagiwara:2011af}	& $-$ & $80\times 10^{-11}$\\
			\hline 
			\hline
		\end{tabular}\end{center}
		\caption{
			Experimental results of the observables combined by the Particle Data Group (PDG) and/or Heavy Flavor Averaging Group
 (HFAG) Collaborations  in Refs.~\cite{Agashe:2014kda}--\cite{Amhis:2014hma}. 
			As for $\overline{\mathcal B} (B^0_q \to \mu^+ \mu^-)$, the combined results from the LHCb and CMS collaborations are shown as
given  in Ref.~\cite{Archilli:2014cla}. Ref.~\cite{Hagiwara:2011af} is used for constraints from $(g-2)_\mu$ data.
		}\label{Tab:ExpResult}
	\end{table}

\section{Parameter space}
\label{sec:param}
\subsection{Tools for theoretical and experimental constraints}
\label{sec:expcons}
For our study, we perform a systematic scan on the parameter space of the 2HDM. The scan is done 
with the help of the public code 2HDMC \citep{Eriksson:2009ws}. 2HDMC calculates the 2HDM spectrum that is consistent with all theoretical constraints such as perturbative unitarity, boundedness from below 
of the scalar potential as well as EW Precision Observables (EWPOs). The code also allows one to calculate 
the decay rates (including Branching Ratios (BRs)) of all Higgs particles. 
We also link the code to Super-Iso \citep{Mahmoudi:2008tp} in order to check 
for consistency with various $B$-physics observables  that we request to be 
 within 2$\sigma$ from the experimental measurements given in Tab. \ref{Tab:ExpResult}.
The direct search constraints from LEP, Tevatron and LHC 
are checked using the the public code HiggsBounds-5 \cite{Bechtle:2013wla}. We also enforce HiggsSignals-2 \cite{Bechtle:2013xfa} constraints from Higgs measurements of LHC data.

\subsection{Production and decay processes}
The aim of this analysis is to find an interpretation within the framework of some 2HDM
to the four-jet excesses observed when re-analysing LEP2 data from ALEPH with Center-of-Mass (CM) energies
over the range $130$ GeV  $\leq \sqrt{s}\leq 208$ GeV \cite{Kile:2017ccn}. As intimated, 
we will identify the CP-even Higgs boson $H$ as the SM-like Higgs state observed by ATLAS and CMS
with $m_H=125$ GeV and scan over the other parameters of the model as described in 
Tab. \ref{tab:parameters}. Since we assume that $H$ is SM-like and data point out that the $HVV$ coupling is almost full strength,
we limit $\sin(\beta-\alpha)$ in the following two ranges: $[-0.25,0]$ and $[-0.6,0.6]$ (in order to have a substantial $HVV$ coupling).

\begin{table}[h!]
	\centering
	{\renewcommand{\arraystretch}{1.2} 
	{\setlength{\tabcolsep}{1.5cm}
	\begin{tabular}{|c | c | c |}
		\hline
		Parameters  & 2HDM-I & 2HDM-III\\
		\hline
		\hline
		$m_h$ (GeV)  & [10,90] & [10,100]\\
		$m_H$ (GeV) & 125.09 & 125.09\\
		$m_A$ (GeV) & [10,90]  & [10,100]\\
		$m_{H^\pm}$ (GeV) & [60, 200] & [50, 200] \\
		$s_{\beta-\alpha}$  & $[-0.25,-0.05]$ & $[-0.6,0.6]$\\
		$\tan\beta$ & [$\frac{-0.98}{s_{\beta-\alpha}}$,$\frac{-1.11}{s_{\beta-\alpha}}$]&[2,50] \\
		$m_{12}^2$ (GeV$^2$)& $m_h^2s_\beta c_\beta$ & $m_h^2s_\beta c_\beta$ \\
		$\lambda_6=\lambda_7$ & 0 & 0\\
		$\chi^{u,d,l}$ & - & [-3,3]\\		
		\hline
	\end{tabular}}}
	\caption{2HDM parameters and their scanned ranges: here, $s_{x}\equiv \sin(x)$ and $c_{x}\equiv \cos(x)$.}
	\label{tab:parameters}
\end{table}

As mentioned,  Ref. \cite{Kile:2017ccn} shows that the four-jet excesses are localised  in the region
$M_1 + M_2 \approx 110 $ GeV (recall that  $M_1$ and $M_2$  are the two di-jet masses). About half of the anomalous events are concentrated in the region $M_1 \approx 80$ GeV and $M_2\approx 30$ GeV while the other half of the events is found when 
$M_1 \approx M_2 \approx 55$ GeV.  In the 2HDM these can come from  the following processes:
\begin{itemize}
\item[1.] $e^+e^- \to h A$ with  $(m_h , m_A)$ or $ (m_A,m_h)$ $\approx (30, 80)$ GeV; 
\item[2.] $e^+e^- \to H^+ H^-$  with $m_{H^\pm}\approx 55$ GeV. 
\end{itemize}
Further, we can also mention that the $ (m_A,m_h)$ $\approx (55, 55)$ GeV solution is not possible in these models.
A viable interpretation of the four-jet excesses will depend of course on the 
cross sections of these processes times the BRs of  $h,A$ and $H^\pm$ into di-jets.

We first note that  LEP constraints  from the Higgs-strahlung process $e^+e^- \to Z^{(*)}h$  can  
 restrict the allowed range of  $\sin^2(\beta-\alpha)$ for a given value of $m_h$, as long as 
 the process is kinematically allowed (even off-shell), which is the case for $\sqrt{s}\approx  m_Z + m_h$ \cite{Abdallah:2004wy}.
 Since the $ZZh$ coupling in the 2HDM suffers from a $\sin^2(\beta-\alpha)$ suppresssion with respect to the SM value,
 and knowing (from LHC data)  that in our scenario $\sin^2(\beta-\alpha)$ must be small, the $e^+e^-\to Z^{(*)}h$ channel is generally compliant with LEP 
experimental constraints. In turn, though, this also means that the corresponding cross section is generally small, even more so when  $\sqrt{s}< m_Z + m_h$, so that the $Z$ boson is off-shell. In fact, we have verified that this channel  production rate is always well below those of the processes in 1--2 above, so that, henceforth, we will neglect it in our analysis. 
For a  low mass $m_h\leq 60$ GeV, with the $h$ state decaying fully into $b\overline{b}$, 
 LEP2 data put a stringent limit on $\sin^2(\beta-\alpha)\leq 0.05$ \cite{Abdallah:2004wy}.
 We further mention that Tevatron also searched for such a light Higgs in the $p\overline{p}\to Vh$ 
production mode~\cite{Aaltonen:2013js}, however, these bounds are much less stringent 
than the LEP ones.
LEP has also searched for a CP-odd scalar $A$ produced in association with a CP-even $h$ in the process 
$e^{+}e^{-} \to h A$~\cite{LEPHiggsWorking:2001ab}. 
This search is complementary to that for $e^{+}e^{-} \to Z^{(*)}h$ 
as the former depends on $\cos^2(\beta-\alpha)$ while the latter depends on $\sin^2(\beta-\alpha)$. 
The null results in  both production modes significantly constrain both $\sin(\beta-\alpha)$ and 
$\cos(\beta-\alpha)$ simultaneously, eliminating large regions of parameter space. If both $h$ and $A$ 
are light at the same time,  such that 
$m_A+m_h < 209$ GeV, then combined direct searches for $e^+e^-\to hA$ (in a variety of final states) rule out a 
significant part of the 2HDM parameter space including the regions which satisfy the alignment limit $\cos(\beta-\alpha)=1$. 
In fact, the constraint from $e^{+}e^{-} \to h A$ is actually  on $\cos^2(\beta-\alpha) \times {\rm BR}(h\to f\bar f)\times {\rm BR}(A\to f'\bar f')$,
where $f\bar f$ and $f'\bar f'$ are possible fermionic decay channels of $h$ and $A$, respectively 
\cite{Abdallah:2004wy,Abbiendi:2000ug,Abbiendi:2004gn}. 
The maximal value $\cos^2(\beta-\alpha)=1$  can in principle exclude large mass regions for $m_{h}$ and 
$m_A$  between 20 and 120 GeV. More precisely,  if $h$ and $A$ decay dominantly (i.e., close to 100\%)  into a $b\overline{b}$ 
pair, then the most stringent limit from $e^{+}e^{-} \to h A$  at LEP2 with CM 
energies $\sqrt{s}=183$ and $187$ GeV is given by \cite{Abbiendi:2000ug}. In practice, 
 for $\cos^2(\beta-\alpha)=1 $, it excludes the mass range $33~{\rm GeV}\leq   m_h, m_A  \leq 78 $ GeV.

As for the $e^+e^-\to H^+H^-$ channel, in the case of only fermionic decays of the charged Higgs boson, there exists a universal (i.e., model independent)  limit on its mass, $m_{H^\pm}>80$ GeV or so, since the $\gamma H^+H^-$ and $ZH^+H^-$ couplings are only  due to the gauge structure of the 2HDM. However, if decays of the type $H^\pm\to W^{\pm *}h$ or (especially) $H^\pm\to W^{\pm *}A$ are allowed, then 
lower $m_{H^\pm}$ values are possible~\cite{Arhrib:2017wmo}. 
 
 Needless to say, the masses and couplings entering the aforementioned $e^+e^-$ processes also affect indirectly the SM-like Higgs data collected at the LHC, not only through the mixing between $H$ and $h$, but also via $H^\pm$ effects in $H\to \gamma\gamma$ and $\gamma Z$ and via $H\to AA$ and $hh$ decays (which would affect the total $H$ width).

\begin{figure}[h!]
	\centering
	\includegraphics[width=0.47\textwidth,height=0.4\textwidth]{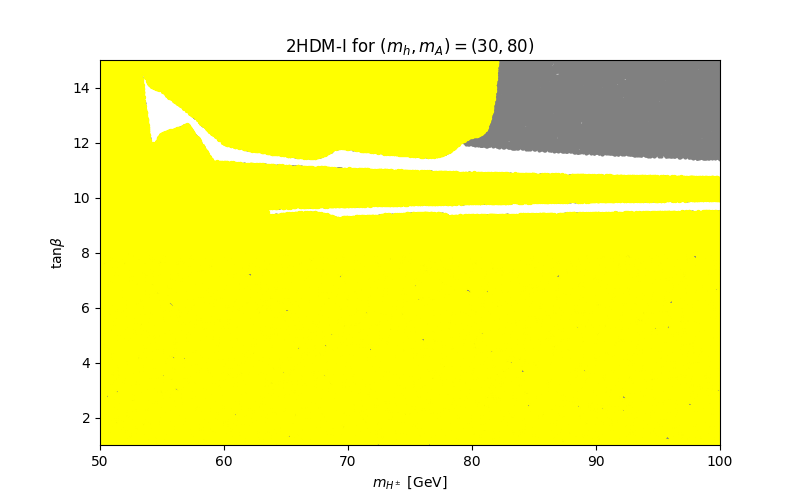} 
	\includegraphics[width=0.47\textwidth,height=0.4\textwidth]{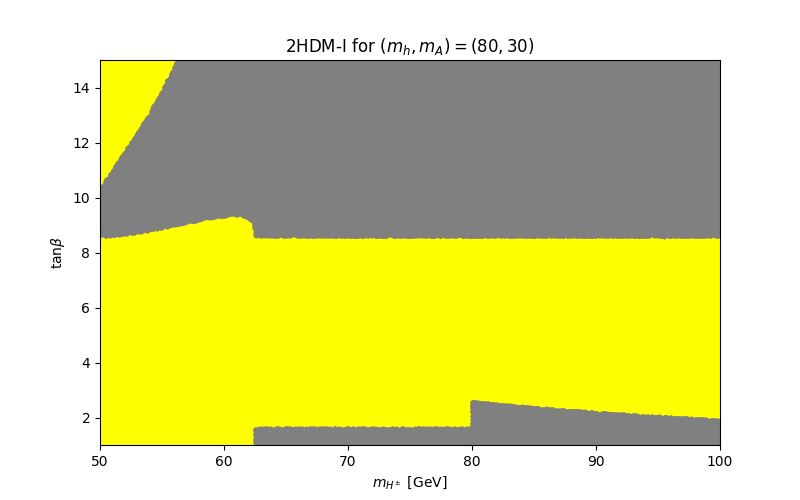}
	\includegraphics[width=0.47\textwidth,height=0.4\textwidth]{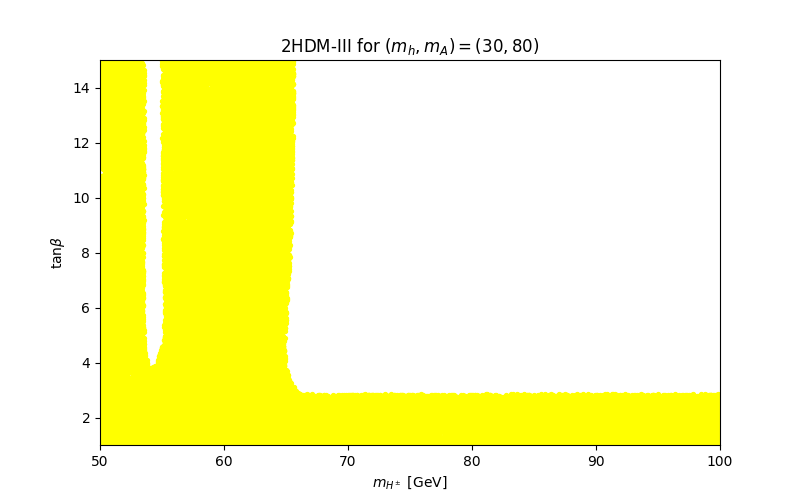} 
	\includegraphics[width=0.47\textwidth,height=0.4\textwidth]{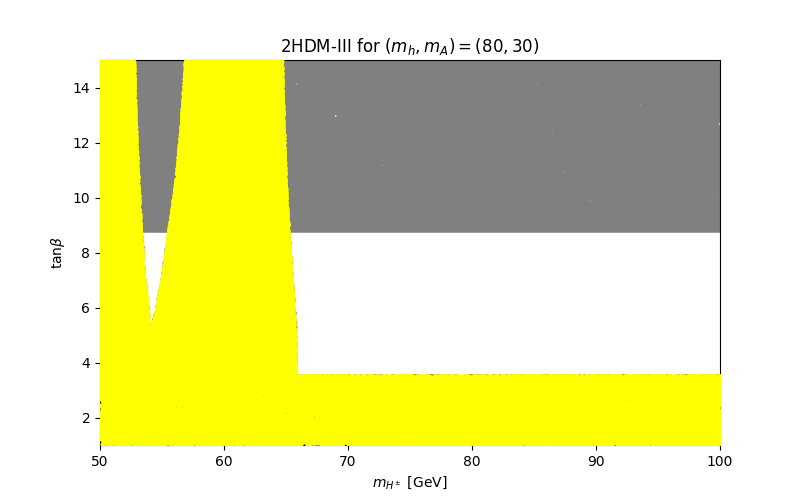}
	\caption{The allowed regions (white) in the $(m_{H^\pm}, \tan\beta)$ plane. 
	Yellow areas are excluded by LEP while gray areas are excluded by LHC Run II.
	Upper panels are for the 2HDM-I and lower ones are for the 2HDM-III.}	
	\label{fig:1a}
\end{figure}

\subsection{Parameter scans}
In order to delineate the impact of LEP and LHC data upon the paremeter spaces of our 2HDM-I and -III  scenarios, we start by  performing a scan on $\tan\beta$, $m_{H^\pm}$  and $\sin(\beta-\alpha)$ in order 
to see what are the 2HDM parameter configurations that are consistent with the $(m_h , m_A)$ or $ (m_A,m_h)$ $\approx (30, 80)$ GeV and 
$(m_{H^\pm}, m_{H^\pm})$ $\approx (55,55)$ GeV 
solutions.  Our results are illustrated in Fig.~\ref{fig:1a} for the 2HDM-I (top) and -III (bottom).
In the left panels, we project the LEP and LHC constraints discussed above (and implemented via HiggsBounds and HiggsSignals)  
onto the allowed regions at 95$\%$ Confidence Level (CL)  in the $(m_{H^\pm},\tan\beta)$ plane for $(m_h , m_A) = (30, 80)$ GeV while in the right panels we illustrate  the case for  $(m_h , m_A) = (80, 30)$ GeV. Regions in yellow are excluded by LEP while those in gray by LHC data, so that only the  white regions are allowed. In the 2HDM-I, one can see that the solution $(m_h , m_A) = (80, 30)$ GeV is totally excluded while the one with $(m_h , m_A) = (30, 80)$ GeV is allowed, though rather restricted, so that  only a tiny region on the $(m_{H^\pm},\tan\beta)$ plane is allowed. This, however, captures simultaneously the $(m_{H^\pm}, m_{H^\pm})$ $\approx (55,55)$ GeV solution. Furthermore, 
as one can see from the lower panels, in the 2HDM-III, we have instead a large area of the $(m_{H^\pm},\tan\beta)$ 
plane which is allowed by  both LEP and LHC data. Not only we have abundant parameter space where both the  $(m_h , m_A) = (30, 80)$ GeV and 
$(m_h , m_A) = (80, 30)$ GeV solutions are possible, but also regions exist where a
charged Higgs boson with $\approx 55$ GeV mass is also allowed.

\begin{figure}[h!]
\centering
\includegraphics[width=0.47\textwidth,height=0.4\textwidth]{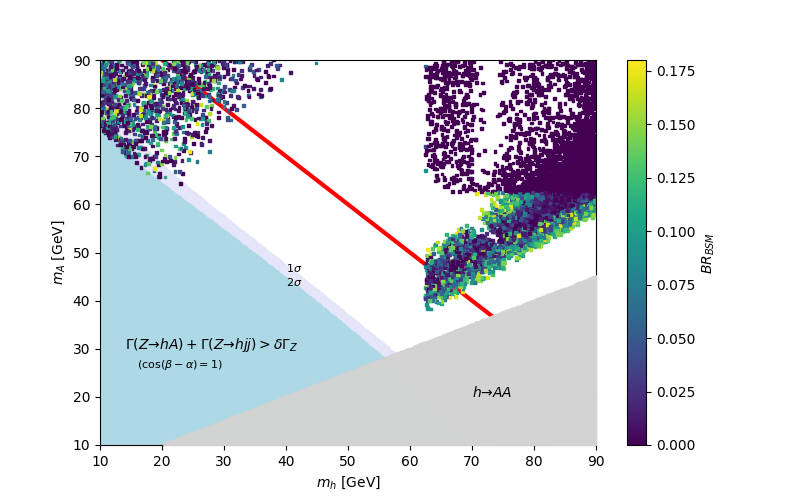} 
\includegraphics[width=0.47\textwidth,height=0.4\textwidth]{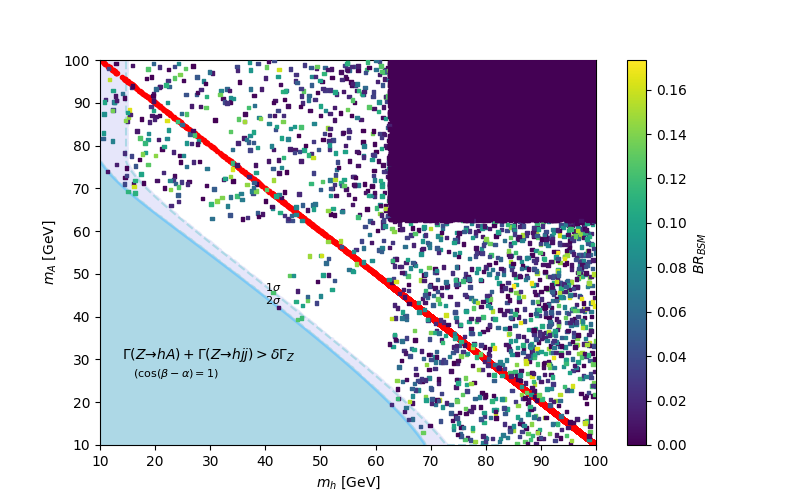}
\caption{The allowed points in the $(m_h, m_A)$ plane with BR$_{\rm BSM}$ indicated in the colored gauge in the 2HDM-I (left) and 2HDM-III (right). The red lines satisfy $m_h+m_A=110$ GeV.  LEP exclusions due to $\Gamma_Z$ and $h\to AA$ data are given by the light blue and light gray shaded areas, respectively.}	
	\label{fig2}
\end{figure}

Finally, Fig. \ref{fig2} shows the actual points generated by our scan (on the left for the 2HDM-I and on the right for the 2HDM-III) that survive all the constraints we discussed, color gauged in terms of BR$_{\rm BSM}$, which is essentially the 2HDM contributions  
to the BR of  invisible $H$ decays, which LHC data presently constrain to be  less than 20\% or so. The points are scattered over the $(m_h,m_A)$ plane, wherein the line corresponding to $m_h+m_A=110$ GeV, capturing the neutral Higgs solutions to the four-jet anomalies, is drawn. We also shade in 
light blue the regions excluded by the $Z$ width measurements, which would be affected by $Z\to hA$ and $Z\to hZ^*$ decays,  and in light gray those excluded by direct $h\to AA$ searches, both performed at LEP. This plot shows that in the 2HDM-I the unavailable neutral Higgs solution, i.e., 
$(m_h,m_A)\approx(80,30)$ GeV, is eliminated by the latter set of data.

\section{Results and Discussions}
\label{sec:results}
In this section, we discuss the yields of the processes $e^+e^-\to hA$ and $e^+e^-\to H^+ H^-$ as predicted by the 2HDM-I and -III at the various LEP energies and luminosities used by the ALEPH analysis and confront these to
the corresponding results. We first estimate the typical efficiency of a four-jet selection on the aforementioned two channels once the Higgs states are allowed to decay hadronically. Then we show that compatibility between ALEPH data and 2HDM-III (but not 2HDM-I) predictions can be achieved on a sizable region of parameter space.

\subsection{Signal selection efficiencies}

The analysis in \cite{Kile:2017ccn} reports a nominal best fit to the data of $N_{80,30} = 121 \pm 33$ and $N_{55,55} = 138 \pm 43$ events for the (80,30) and (55,55) GeV excesses, respectively. They find that the existence of the excesses is robust with respect to choice of MC event generator, jet-clustering algorithm and MC reweighting procedure, but that the best fit number of excess events varies with these choices.  In particular, the best fit to $N_{80,30}$ went as low as $92\pm 25$ and the best fit to $N_{55,55}$ went as high as $203\pm 56$ events. Furthermore, the results presented in \cite{Kile:2017ccn} are at preselection level, with no analysis cuts to enhance the resonance.  While complete comparison would require a full MC analysis including detector simulation and appropriate jet reconstruction algorithms, given the loose selection and the variation in the size of the best fit, here we use parton-level cross sections and BRs to estimate the number of signal events corresponding to the 2HDMs considered here\footnote{To truly explain the excess as originating from two resonances as we propose here, the kinematic distributions of the excess would be needed, yet they are not currently available.}.  

To improve our estimate, we consider that the analysis requires reconstruction of two resonances from four-jets, which leads to combinatoric issues which can reduce the number of signal events in the excess regions. In particular, the jet pairing is chosen as follows. 
\begin{itemize}
\item The invariant mass difference between the two jet pairs is minimized.
\item $M_1$ is assigned to the pair containing the highest $p_T$ jet.
\end{itemize}

To estimate the effect of this selection, we simulate $e^+ e^-\to hA \to 4j$ events at all CM energies present in the data set and masses of 80 and 30 GeV for the Higgs states using MadGraph5\_aMC@NLO\cite{Alwall:2014hca}. 
This is done at parton level with negligible widths for $h$ and $A$, and we calculate the fraction of events for which the jets are correctly paired and for which the 80 GeV resonance contains the highest $p_T$ jet.  We take these as the selection efficiencies, reported in Tab.~\ref{tableefficiency}.  While our scan allows our resonances to deviate slightly from (80,30), we find the variation of efficiencies to be too small to qualitatively change any findings, so for simplicity we apply the values int Tab.~\ref{tableefficiency} to all points in the scan.  We further note that at the level of calculation presented here, with all jets perfectly reconstructed, the correct mass pairing will always be chosen for the (55,55) solution, so we do not apply any efficiency to the $e^+ e^-\to H^+ H^-$ cross sections.

\begin{table}[h!]
\centering
\begin{tabular}{c|c|c}
$\sqrt{s}$ [GeV] & ALEPH data [pb$^{-1}$] & Efficiency\\ \hline
130.0 & 3.30 & 0.32\\
130.3 & 2.88 & 0.32\\
136.0 & 3.50 & 0.26\\
136.3 & 2.86 & 0.26\\
140.0 & 0.05 & 0.25\\
161.3 & 11.08 & 0.22\\
164.5 & 0.04 & 0.22\\
170.3 & 1.11 & 0.22\\
172.3 & 9.54 & 0.23\\
182.6 & 59.37 & 0.23\\
188.6 & 177.08 & 0.24\\
191.6 & 29.01 & 0.24\\
195.5 & 82.62 & 0.24\\
199.5 & 87.85 & 0.24\\
201.6 & 42.14 & 0.25\\
204.9 & 84.03 & 0.25\\
206.5 & 130.59 & 0.25\\
208.0 & 7.73 & 0.25
\end{tabular}
\caption{Combinatoric efficiencies for the (80,30) GeV solution for the energies used in the LEP2 analysis. }
	\label{tableefficiency}
\end{table}

\subsection{The four-jet excesses in the 2HDM-I and -III}

To address the excesses in the context of the 2HDM, we perform refined scans of the relevant parameter space: $m_h$(or $m_A$) is confined to $(25,35)$~GeV, $m_A$(or $m_h$) is confined to $(75,85)$~GeV, and $m_{H^\pm}$ must be in the range $(50,60)$~GeV.  The other parameters are as in Tab.~\ref{tab:parameters}.
For each point in our scans, we calculate expected number of excess events in each region as

\begin{equation}
N = \sum_{\{\sqrt{s}\}} \mathcal{L}(\sqrt{s}) \times \sigma_{4j}(\sqrt{s}) \times \epsilon(\sqrt{s}).
\end{equation}
Here the sum is over the CM energies used in the experimental analysis, as shown in the first column of Tab.~\ref{tableefficiency}, and $\sigma_{4j}$ is the cross section $\sigma(e^+ e^- \to h A \to 4j)$  for the (80,30) GeV excess whereas $\sigma(e^+ e^- \to H^+ H^- \to 4j)$ is for the (55, 55) GeV excess\footnote{While other channels involving states from the Higgs sector could also produce final states with four jets, we do not find any of them to contribute significantly to the excess regions, so we do not consider them here.}. The luminosities, $\mathcal{L}$, and the (80,30) efficiencies, $\epsilon$, at each energy are given in the second and third columns of Tab.~\ref{tableefficiency}, respectively. As described in the previous section, the efficiencies for the (55,55) GeV excess are taken to be unity.  As discussed above, a point should produce $\mathcal{O}$(100) events for each excess region to be considered a viable explanation.

\begin{figure}[h!]
\centering
\includegraphics[width=0.47\textwidth,height=0.4\textwidth]{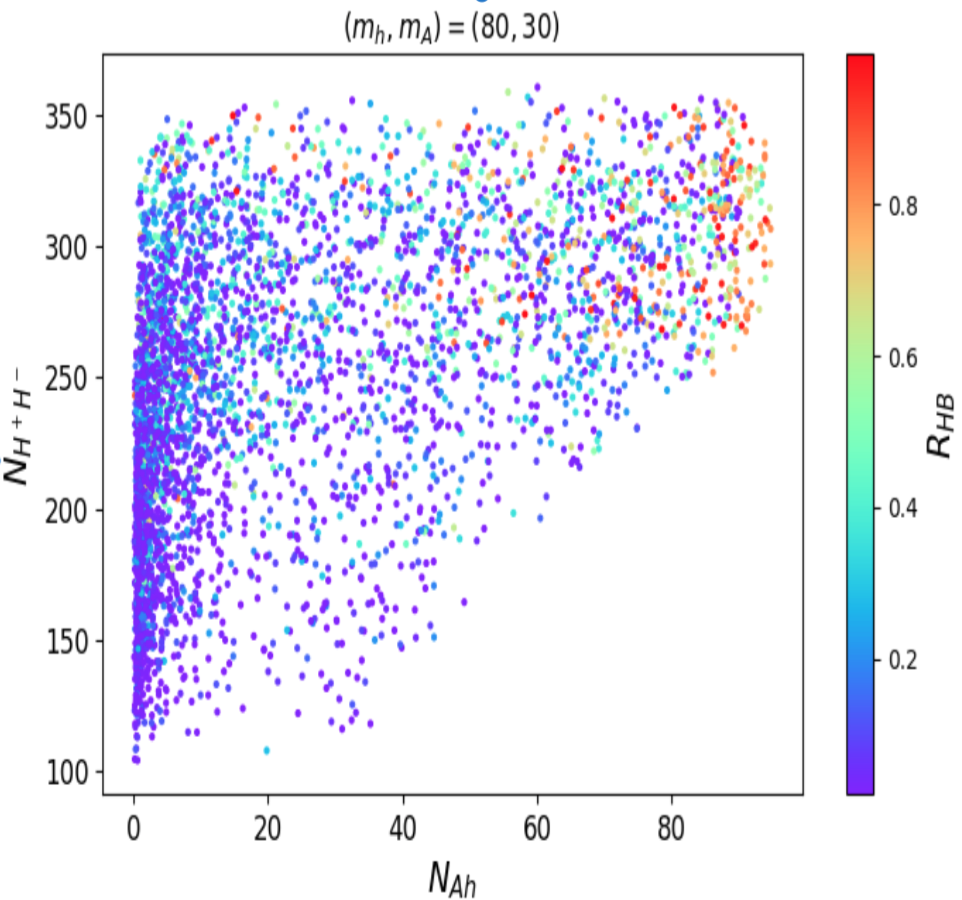} 
\includegraphics[width=0.47\textwidth,height=0.4\textwidth]{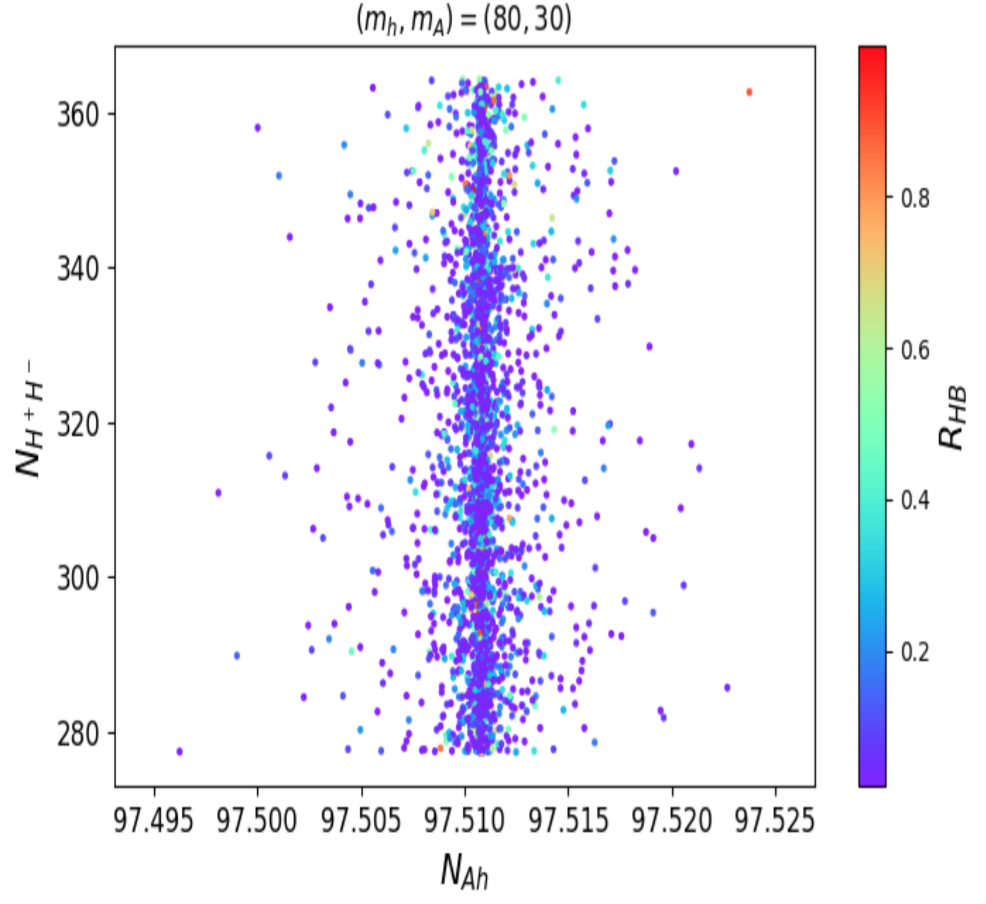}
\caption{Correlation between $N_{hA}$ and $N_{H^\pm H^\mp}$ with $R_{HB}$ indicated in the right palette: on the left panel we consider actual BRs while in the right panel we assume  BR$(\Phi\to jj) = 100\%$ ($\Phi = h, A, H^\pm$) . }	
	\label{fig:Nevents}
\end{figure}

We find that the points from the 2HDM-I scan do not have large enough cross sections to fully account for the excesses (with maximum values of 23 and 86 events for the (80,30) and (55,55) solutions, respectively), so the remainder of this paper will deal entirely with the 2HDM-III scenarios.  Fig.~\ref{fig:Nevents} shows the correlation between $N_{Ah}$ and $N_{H^+H^-}$ for the case $m_h \sim 80~\textrm{GeV}$, $m_A \sim 30~\textrm{GeV}$.  It is clear in the left panel that the $H^+H^-$ channel readily produces $\mathcal{O}(100)$ events.  Because of the combinatoric inefficiencies in the (80, 30) channel, the number of four-jet events from the $hA$ channel is smaller, but many points still appear consistent with the observed excess at this level.  The color map in Fig.~\ref{fig:Nevents} gives the ratio $R_{HB}$ of the theoretical signal to experimental limit across all analyses included in HiggsBounds (i.e. points with $R_{HB} > 1$ are excluded), indicating that the set of points of interest contains scenarios which were nearly excluded by past searches as well as those which lie safely away from the limits considered.

In the right panel of Fig.~\ref{fig:Nevents}, the BRs of all Higgs states to jets are set to unity. The resulting region is significantly more compact, indicating that the number of excess events is driven mainly by the decays of the Higgs states and that there is little variation in the production cross sections, especially in the $hA$ channel, where the important $ZhA$ coupling is pushed towards its maximal value by requiring a SM-like 125 GeV state.  The maximal values in this plot are also not much larger than those for the left panel.  For the $hA$ channel, this demonstrates that the scan nearly saturates the theoretical upper limit. We have also considered the case where $m_h \sim 30~\textrm{GeV}$, $m_A \sim 80~\textrm{GeV}$.  This scenario produces similar results for the $H^+H^-$ channel, but does not produce the required rate in the $hA$ channel, with a maximum excess of about 40 events only.

While the effects of a full experimental simulation and analysis would likely somewhat diminish the approximate results shown here, given the loose selection and lack of analysis cuts in finding the excess, a selection of points found here could provide a plausible explanation.

\section{Conclusions}
\label{sec:summa}
In this paper, we have shown how some excesses recently isolated by a re-analysis of ALEPH four-jet data collected at LEP2 can potentially be ascribed to the production and decay of Higgs boson pairs in the 2HDM-III, wherein the relevant production processes are $e^+e^-\to hA$ and $e^+e^-\to H^+ H^-$ followed by $h, A$ and $H^\pm$ hadronic decays in all possible channels. Under the assumption that the SM-like Higgs boson discovered at the LHC in 2012 is the heaviest CP-even Higgs state, $H$, of this construct, we have isolated sizable regions of the parameter space of this new physics scenario wherein the required number of signal events is produced, thus explaining the exesses seen in the di-jet mass combinations (80, 30) GeV (via $hA$ intermediate states) and (55, 55) GeV (via $H^+H^-$ inteermediate states). This has been eventually done after implementing all available experimental constraints from both collider and non-collider experiments and in presence of theoretical conditions of self-consistency of the 2HDM versions that we proposed. 

Although our kinematical analysis was not refined as it could be, as we have not performed a full MC analysis in the  presence of parton shower, hadronization and detector effects through a proper jet-clustering based reconstruction of four-jet samples, we are confident that our results are solid enough so as to call for a more thorough experimental investigation of the 2HDM-III dynamics advocated here as a possible theoretical explanation of puzzling LEP2 results.  In this connection, we finally highlighted the fact that the excesses discussed here for the 2HDM-III occur in regions of their parameter space that can also be tested by the LHC with present and upcoming data, as shown in previous publications of ours. 

\section*{Acknowledgments}
SM is supported in part through the NExT Institute and the  STFC  CG  ST/L000296/1. 
All authors acknowledge   financial  support  from    the  H2020-
MSCA-RISE-2014 grant no.  645722 (NonMinimalHiggs).



\end{document}